\documentstyle[12pt,psfig]{article}

\title{Rotating Rayleigh-B\'enard Convection: Aspect Ratio Dependence of the
Initial Bifurcations}

\author{Li Ning and Robert E. Ecke \\
{\small {\em Physics Division and Center for Nonlinear Studies}} \\
{\small {\em Los Alamos National Laboratory, Los Alamos, NM 87545}}}
\begin{document}

\maketitle

\begin{abstract}

The initial bifurcations in rotating Rayleigh-B\'enard convection are studied
in the range of
dimensionless rotation rate $0 < \Omega < 2150$ for an aspect-ratio-2.5
cylindrical cell. We
used simultaneous optical shadowgraph, heat transport and local temperature
measurements to
determine the stability and characteristics of the azimuthally-periodic wall
convection state. We also show that the second transition corresponds to the
onset
of bulk convection. Our results for critical Rayleigh numbers, precession
frequencies
and critical mode numbers agree well with theoretical results. The dynamics of
the
wall convection state can be described by a complex Ginzburg-Landau amplitude
equation.

\end{abstract}

\section{Introduction}

	Studying the effects of external rotation on thermal convection has attracted
significant experimental and theoretical interest
\cite{Ch61,CB79,Ro69,ZES91}. Because of its general occurrence in
geophysical and oceanic flows, it is important to understand how the Coriolis
force influences the structure and transport properties of thermal convection.
Rotating thermal convection also provides a system in which to study
hydrodynamic
instabilities \cite{Ro69,HH72,BC83}, pattern formation and spatio-temporal
chaos in
nonlinear dynamical systems \cite{ZES91,KL69}, and transitions to turbulence
\cite{Bu81,ND81}.

	Rayleigh-B\'enard convection consists of a fluid layer confined between two
horizontal plates and heated from below. External rotation is imposed by
rotating the convection cell about a vertical axis. Most theoretical
studies of this system have assumed a laterally infinite cell geometry
\cite{Ch61}. In
experiments, however, there are always side walls. Moreover, to achieve
sufficient laboratory control over the experimental parameters and obtain large
dimensionless rotation rates, the horizontal dimensions of the convection cells
are usually comparable to their vertical depths. This raises the question
whether the theoretical predictions for infinitely extended systems are
relevant
for real experiments. It has indeed been found that the presence of lateral
boundaries change significantly some aspects of thermal convection in the
presence of
rotation\cite{Ro69,ZES91,BC83}. In this paper, we present
experimental results that illustrate clearly the influence of the lateral
boundaries. We confine ourselves to cylindrical convection cells and
characterize
the geometry by the aspect ratio which is the ratio of the radius to the depth.

	There have been several experimental and theoretical investigations on
rotating
Rayleigh-B\'enard convection. One of the early quantitative experiments by
Rossby\cite{Ro69} showed significant deviation from the predictions of theories
for a
laterally infinite system.
His measurement of heat transport showed a slow increase over the pure
conduction state at temperature differences much smaller than the expected
onset of
convection for an infinitely extended system. There have been many explanations
offered
to remedy this discrepancy. The first one that correctly attributed the
increase in
thermal transport to a side wall convection state was proposed by Buell and
Catton\cite{BC83}. They investigated the stability of the conduction state in a
laterally finite cylinder in the presence of rotation. Contrary to conventional
wisdom
that side walls suppress thermal convection, they found that for a large enough
rotation rate, some side wall convection states have a critical control
parameter that is much lower than the value for the laterally infinite case.
These
non-axisymmetric wall states are azimuthally periodic and have maximum
amplitude
near the side wall. Zhong {\it et al}\cite{ZES91} later reported experimental
confirmation of the wall convection states in a unit aspect ratio cell.
They observed, however, that these asymmetric states precessed slowly counter
to the
direction of rotation. In addition to the first bifurcation to the wall states,
there was a second bifurcation when the central region became filled with
vortex-like structures, accompanied by an increase in the thermal transport and
in the thermal noise measured in the bottom plate. Goldstein {\it et
al}\cite{GKMN92} carried out a linear stability analysis which allowed for the
experimentally observed precession and found good agreement with the
experimental
results. Ecke {\it et al}\cite{EZK92}
explained the traveling wall states by a broken azimuthal
reflection symmetry and  predicted that the precession frequency would go
linearly to
zero as the rotation rate went to zero\cite{EZK92}.

	Our experiment, using an aspect ratio $2.5$ cell, was designed to explore the
aspect ratio dependence of the initial bifurcation from the conduction state to
the wall convection state.
We chose to increase the aspect ratio only moderately relative to the previous
cell
of Zhong et al \cite{ZES91} because the range of accessible dimensionless
rotation rates
(limited by the condition that centrifugal effects be small) for the larger
aspect ratio cell
overlapped a large fraction of the parameter range used in the previous work.
Furthermore, although the physical aspect ratio is considered small for
nonrotating
convection, rotation increases the effective aspect ratio significantly. In
addition, since the side-wall state forms around the azimuth, the number of
structures that fit in that geometry is roughly the ratio of the perimeter to
twice
the depth. For $\Gamma$ = 2.5 this ratio is 7.9. Thus, provided the
dimensionless
rotation rate is sufficiently large, one can consider the conditions to be in
the
moderate to large aspect ratio limit and our measurements presented here
together
with the previous work for $\Gamma$ = 1 suffice to characterize the aspect
ratio
dependence of the observed bifurcations. We discuss this in more detail later.

In this work we present measurements of the critical Rayleigh numbers and
critical
wavenumbers for the initial transition to a side-wall traveling wave state and
the
second transition to bulk convection. We find that the traveling wave state can
be
described quantitatively by a one-dimensional complex Ginzburg-Landau (CGL)
equation.
In addition we find that the traveling wave state does not depend on the
overall
geometry of the convection cell. Rather it exists on any wall regardless of the
magnitude or sign of the wall curvature. Further, the conductivity of the wall
is
shown to strongly influence the critical Rayleigh number for the bifurcation to
traveling waves.
Finally, the second transition to bulk convection is shown to occur as a
spatially seperate
transition owing to the negligible amplitude of the wall state in the center of
the
cell.

\section{Experiment}

The apparatus used in these convection studies was descibed in detail by Zhong
{\it et al}\cite{ZES91}. The bottom plate of the convection cell was a
nickel-plated, mirror-polished, copper plate with a thermofoil heater attached
to
the back. The top plate was a $1/8\ inch$ thick, optical quality sapphire
window. The
side wall was made of 0.31 cm thick plexiglasss, defining an active circular
area with a
radius $r = 5.00\ cm$. The depth $d$ of the cell was $2.00\ cm$, giving an
aspect ratio
$\Gamma = r/d = 2.5$. We used an afocal optical shadowgraph technique to
visualize the convective flow field. Because of the depth of the cell (hence a
very small temperature gradient), the shadowgraph method was not sensitive
enough
to observe the flow near the first bifurcation, especially when the rotation
rate
was small. For local temperature measurement, two thermistors were embedded at
half-height positions in the side wall with known angular separation. Since the
wall states traveled along the circumference, the thermistors measured the
temperature field sweeping by. From these data, the amplitude,
frequency and mode number of the traveling wave were extracted. The heating was
provided by applying a constant voltage to the bottom heater, and a
circulating,
temperature-regulated water bath outside the top sapphire window provided
cooling and constant top temperature to better than $0.5\ mK$ rms. Since the
convection cell rotates relative to the regulated water bath, the top-plate
temperature is azimuthally very uniform on the time scale relevant to the
dynamics
of the system. Thermistors
embedded in the bottom plate and near the top plate measured the respective
temperatures to a precision better than $0.1\ mK$. Rotation was provided with a
stepper motor with rotation frequency constant to $\pm 0.1\%$. Because the
shadowgraph optics was at rest in the laboratory, a shaft encoder strobed the
video camera at each revolution of the cell. The image digitized via a digital
frame grabber was thus synchronized with the rotating cell. Since the dynamics
of
the convection patterns were slow compared to the rotation period, we did not
lose
any detailed information by recording the images once per revolution.

The control parameters of the experiment are Rayleigh number $R$ and
dimensionless rotation rate $\Omega$, defined as $R = \alpha g d^3 \Delta
T/\kappa \nu$ and $\Omega = \Omega_D d^2/\nu$, where $\alpha$, $\kappa$ and
$\nu$
are the thermal expansion coefficient, thermal diffusivity and kinematic
viscosity of the working fluid, $g$ is the gravitational acceleration, $\Delta
T$ is the bottom-top temperature difference and $\Omega_D$ is the angular
rotation frequency. Another parameter that characterizes the fluid
is the Prandtl number $P=\nu/\kappa$, which is about $6.4$ for our choice of
water at a
mean temperature of about $23.5^\circ C$. A reduced bifurcation parameter is
defined as $\epsilon = (R - R_c(\Omega))/R_c(\Omega)$, where
$R_c(\Omega)$ is the critical Rayleigh number for the onset of the first
convection state at a particular $\Omega$. As we will dicuss below, the first
bifurcation can be to different states depending on $\Omega$. Therefore, we
denote
the critical Rayleigh number for the side-wall state as $R_{cs}$ and that for
the
onset of bulk convection as $R_{cb}$.
The effect of the centrifugal force can
be characterized by the ratio of the centrifugal-to-gravitational force,
$\Omega_D^2 r/g$, which is less than $0.12$ for our maximum rotation rate
$\Omega_D \approx$ 4.9 rad/sec ($\Omega \approx 2145$).

The experimental protocol we used was as follows: $\Omega_D$ was held constant
for the desired $\Omega$ at $\Delta T_c$; the heating was quasistatically
ramped
to each measurement point. After several vertical thermal diffusion times
$\tau_\kappa = d^2/\kappa \approx 2760\ sec$, measurement of temperatures were
made spanning at least one $\tau_\kappa$. Shadowgraph images were recorded at
fixed time intervals during the temperature measurements. We followed the
dynamics
at several parameter values for longer than $40\tau_\kappa$, and no
change of the convection states was found for the parameter range studied.
There
is one exception concerning the stability of the wall state which will
be discussed later.

For determining the onset of the wall convection state, two different
measurements were used for large enough values of $\Omega$ so that the
transition
to the wall state was the first bifurcation. For smaller $\Omega$ only the
second
method was used. The first was the measurement of heat transport,
expressed in Nusselt number $Nu$ which is the ratio of the heat
transported by the convection state to the heat transported by conduction
alone.
In the conduction state $Nu = 1$ whereas just above the onset of
convection, $Nu$ increases linearly with increasing $R$. The intercept
of the linear section of $Nu$ versus $R$ with $Nu = 1$ determines $R_{cs}$. The
second method used the temperature measurements from the side wall thermistors
$T_{md1}$ and $T_{md2}$. By fitting the oscillatory signals to a sine function,
we
extracted the amplitude $\delta T$, which rises above the onset of the side
wall
state as $(R - R_{cs}(\Omega))^{1/2}$. The two methods yielded values of $R_c$
that agreed
to within experimental uncertainty, usually better than $0.2\%$. The second
method
also gave us the frequency and mode number of the traveling wave.

The second bifurcation was determined by the Nusselt number
measurement. We did, however, find two other characteristics of the bulk
convection state that could be used to determine the onset, namely, the thermal
noise in the bottom temperature and the amplitude of the shadowgraph images.
These
results will be discussed in detail later.

\section{Experimental Results and Comparison with Theories}

We present our experimental results on an aspect ratio $2.5$ cell. When
appropriate for showing variation as a function of aspect ratio, we also
include
results for the unit aspect ratio cell from Zhong {\it et al}\cite{ZES91}.
Comparison with theories are made whenever theoretical results are available.

\subsection{Asymmetric wall convection states}

As the temperature difference between the bottom and top plates increases at
fixed $\Omega$ ($\Omega$ actually changes slightly in the experiment since we
held
$\Omega_D$ constant; typically, $\Omega$ varies less than $0.4\%$ for the
range of Rayleigh numbers we used), the first bifurcation from
the quiescent conduction state is to an azimuthally periodic wall convection
state
for moderate to high rotation rates. This convection state precesses uniformly
in
the rotating frame, counter to the direction of external rotation and with
maximum
amplitude near the side wall. The amplitude grows above the onset as
$\epsilon^{1/2}$ and the frequency $\omega_\kappa$ changes linearly with
$\epsilon$ and has a finite value at onset.
This indicates that the bifurcation is a forward Hopf bifurcation.
The unique direction of precession arises from the weak breaking of azimuthal
reflection symmetry by the external rotation (Ecke {\it et al})\cite{EZK92}.
The $\epsilon$
dependence of the amplitude and the frequency of the wall convection state is
adequately characterized in Zhong {\it et al}\cite{ZES91}. We concentrate on
the $\Omega$ dependence of the critical frequencies and mode numbers, and in
particular, on their behaviors as $\Omega$ goes to zero or becomes large.
We also discuss the linear stabilities for different mode-number wall
convection
states.

\begin{figure}
\centerline{\psfig{file=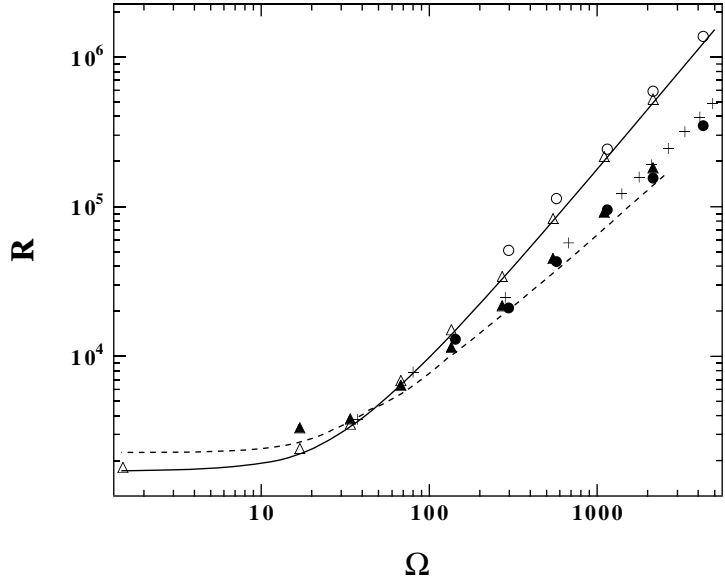,height=2.9in}}
\caption{Parameter-space diagram of rotating Rayleigh-B\'enard convection in
small aspect ratio cylindrical containers. Measured
bifurcation points from conduction to wall convection ($\triangle$) and bulk
convection
states ($\triangle$) for an aspect ratio 2.5 cell and
corresponding bifurcation points ($\bullet$ and $\circ$) measured in a unit
aspect ratio
cell\protect{\cite{ZES91}} are shown. Rossby's results (+) for  aspect
ratios 1.4-5\protect{\cite{Ro69}}, Chandrasekhar's linear stability calculation
for a laterally infinite system (---), and Goldstein {\it et al}'s
result (- - - -) for a unit
aspect ratio cell with insulating side walls\protect{\cite{GKMN92}} are also
shown for
comparison.}
\label{fig:pdiag}
\end{figure}

We have measured the critical Rayleigh number of the wall convection states in
a wide
range of $\Omega$. The results are displayed in Figure \ref{fig:pdiag}. The
dashed line
is from the calculations of Goldstein {\it et al} for perfectly insulating side
walls\cite{GKMN92}, corresponding reasonably well with the experimental
boundary
conditions (plexiglass side wall). The agreement between
experiments and theory is very satisfactory. The results of Zhong {\it et al}
for
$\Gamma = 1$ are also shown in Figure \ref{fig:pdiag}. For completeness, we
have
included  
experimental data of Rossby\cite{Ro69} using cylindrical convection cells with
a
range of aspect ratios from 1.4 to 5.0. There is not much variation in $R_{cs}$
among
these  cell geometries for $\Omega >$ 100.
We believe that the wall-state bifurcation curve is quite
insensitive to further increases in $\Gamma$. The other bifurcation line and
the
cross-over of the two bifurcation lines in \{$R, \Omega$\} parameter space will
be
discussed later.

\begin{figure}
\centerline{\psfig{file=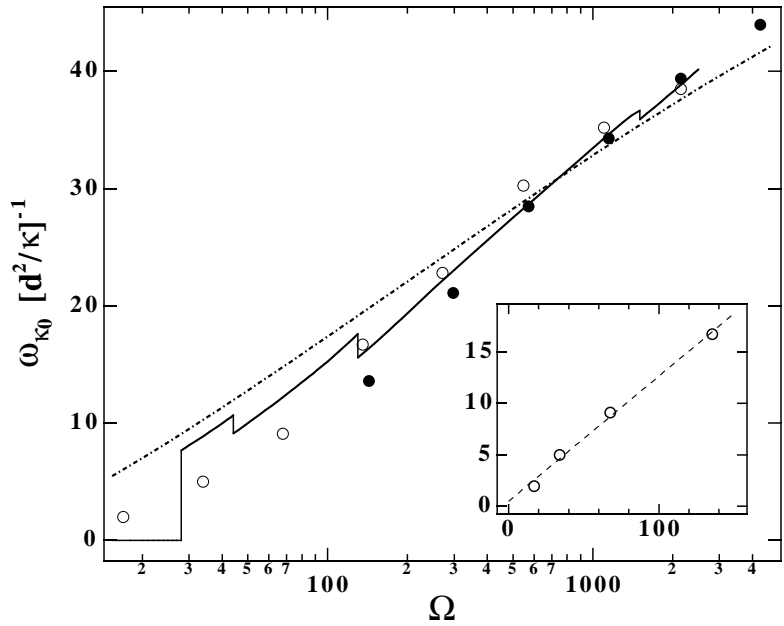,height=3.0in}}
\caption{Comparison of experimental and theoretical results on the
precession frequencies of the wall convection states. Data for aspect ratio 2.5
($\circ$) and 1 ($\bullet$) cells, Goldstein {\it et al}'s results (---) for a
unit
aspect ratio cell, and the results of Kuo and Cross (-$\cdot$-$\cdot$-)
for a planar wall state.
The inset is a linear plot of the frequencies at small rotation
rate and the dashed line is a fit to the data.}
\label{fig:freq}
\end{figure}

	Figure \ref{fig:freq} shows the measured precession frequency at
onset $\omega_{\kappa 0}$ scaled by $\tau_\kappa$ for the wall states versus
$\Omega$ in
a semi-log plot. The inset shows the linear dependence of the critical
frequency on
$\Omega$ for $\Omega < 150$. The
large $\Omega$ behavior of $\omega_{\kappa 0}$ suggests a nearly logarithmic
dependence on $\Omega$. Notice that for high $\Omega$s, the
frequencies for aspect ratios $1$ and $2.5$ cells approach each other. The
solid
line in the figure comes from a recent calculation by Goldstein {\it et al} for
a unit aspect ratio geometry\cite{GKMN92}. Their calculation also  shows that
for
moderately high $\Omega$, the change in frequency with increasing aspect ratio
is
small for $\Gamma > 1$. The dash-dotted curve is the result for infinite aspect
ratio
cells with insulating
rigid side wall and free-free top-bottom boundary conditions from Kuo and
Cross\cite{KC93}.

\begin{figure}
\centerline{\psfig{file=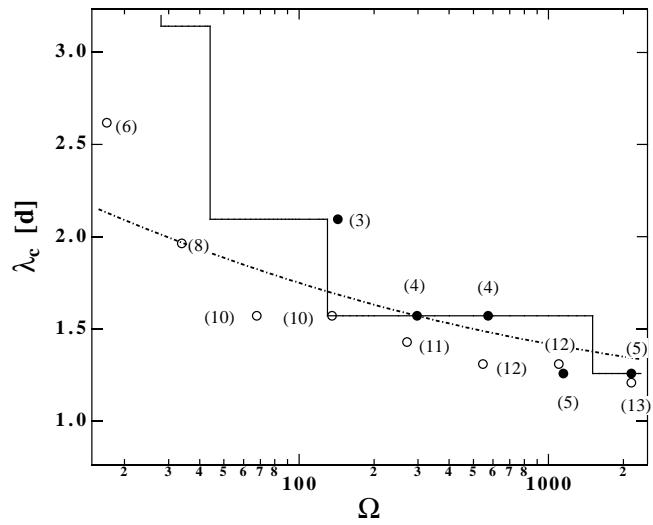,height=3.0in}}
\caption{The critical wavelength for wall convection states with
$\Gamma$=2.5 ($\circ$) and $\Gamma$=1 ($\bullet$). The
bracketed numbers are the corresponding mode numbers. The solid line is from
Goldstein
{\it et al} for a unit aspect ratio cell and the dash-dotted curve is from Kuo
and Cross
for a planar wall state.}
\label{fig:wave}
\end{figure}

The critical mode numbers $m_c$ of the wall states were measured by
quasistatically increasing $\Delta T$ from below. To scale out the aspect ratio
dependence and identify the trend as both $\Gamma$ and $\Omega$ increase, we
convert the mode numbers to dimensionless azimuthal wavelengths scaled by the
depth d,  so that $\lambda_c = 2\pi \Gamma/m_c$. In Table~\ref{precTable} we
show data for the onset of the side-wall state denoted $R_{cs}$,
$\omega_{\kappa 0}$, $m_c$, $\lambda_c$, and the critical Rayleigh
number $R_{cb}$ and the critical wavenumber $k_{cb}$ for bulk convection
for the full range of
\begin{table}
  \begin{center}
	\begin{tabular}{ccccccc}
	 $\Omega$ & $R_{cs}$ & $\omega_{\kappa 0}$ & $m_c$ & $\lambda_c$ & $R_{cb}$ &
	 $k_{cb}$\\
	 & 10$^4$ & & & & 10$^4$ & \\
		0   & ---  & --- & --- & --- & 0.18 & ---\\
		16.93 & 0.33 & 1.98 & 6 & 2.62 & 0.24 & ---\\
		33.86 & 0.38 & 5.0  & 8 & 1.96  & 0.35 & ---\\
		67.70 & 0.64 & 9.1 & 10 & 1.57 & 0.69 & ---\\
		135.5 & 1.14 & 16.7 & 10 & 1.57 & 1.5 & 7.0\\
		271.0 & 2.16 & 22.8 & 11 & 1.43 & 3.4 & 7.9\\
		545.8 & 4.51 & 30.25 & 12 & 1.31 & 8.13 & 11.3\\
		1091  & 9.15 & 35.2  & 12 & 1.31 & 21.5 & 13.9\\
		2145 & 18.1  & 38.5  & 13 & 1.21 & 52.2 & 18.5\\
	\end{tabular}
  \end{center}
\caption{Values of $\Omega$, $R_{cs}$, $\omega_{\kappa 0}$, $m_c$, $\lambda_c$,
$R_{cb}$, and
$k_{cb}$.}
\label{precTable}
\end{table}
$\Omega$ values we explored for the $\Gamma$=2.5 convection cell.
Figure \ref{fig:wave} shows $\lambda_c$
versus $\Omega$ for $\Gamma$ of $1$ and $2.5$. The numbers in parentheseses
near the
data points are the corresponding mode numbers. The stepped solid line is from
Goldstein {\it et al}\cite{GKMN92} for $\Gamma = 1$. The large $\Omega$
behavior
of $\lambda_c$ suggests that it approaches a value near 1 or, in other words,
that the
critical mode number approaches a finite constant. This implies that the depth
of
the convection cell is setting the spatial scale of this wall state.

We can understand the behavior of $m_c(\Omega)$ in the following manner: using
a
physical argument by Manneville\cite{Ma90}, we know that to minimize the
dissipation losses due to vertical and horizontal shear, the convection
rolls should have similar width and height. In the case of wall convection in
the presence of rotation, since 1) the Coriolis force only serves to skew the
particle trajectories without providing kinetic energy, and 2) there is
additional
viscous damping in the vertical direction because the wall states lie close to
the side wall, the convection structure should have a width comparable to
but narrower than if it was not near a rigid wall. This leads to a
critical wavelength in the streamline plane less than $2$ for the wall states.
In
the bulk convection, external rotation gives rise to increasing apparent
wavenumbers(measured as the distance between the periodic structures) as
$\Omega$
increases\cite{Ve59}. In our experiment, since the particle trajectories near
the
side wall have to be nearly parallel to the wall, the measured
wavelength is in the streamline plane. From the radial basis functions for the
velocity and temperature perturbation field, which are $m$th order Bessel
functions $J_m(k r)$, we can estimate the width of the convection structure in
radial direction to be $\delta r \approx X_m/k$, where $X_m \approx 3$ is the
width of
an oscillation period for Bessel functions, and $k$ has to satisfy a dispersion
relation\cite{GKMN92}. Since k increases with $\Omega$, the radial width of the
wall states decreases as
$\Omega$ increases. We can reasonably argue that the wavelength at the middle
of
this width is $O(d)$, independent of rotation. Then $m_c(\Omega) = 2\pi (r -
\delta r/2)/\lambda \approx 2\pi(\Gamma - X_m/kd)$, which gives for
$\Gamma$=2.5 an
asymptotic critical mode of m$_c \approx$ 16. This explains why the critical
mode
number increases with $\Omega$ but saturates at about $2\pi\Gamma$.

In addition to the quantitative measurements presented above we performed a
series
of qualitative investigations to test the relevance of different cell
geometries
and side wall boundary conditions. In one arrangement a circular plexiglass
wall of
half the radius of the convection cell was place concentrically inside the
cell,
thus forming an annulus of width to depth ratio 1.2. Upon increasing R above
$R_{cs}$, wall states formed on all sides of the walls. In the annulus there
was a
traveling wave (TW) on the outer rim that traveled opposite to the direction of
rotation while on the inner rim the TW traveled in the same direction as the
rotation direction. In another arrangement, horizontally straight plexiglass
walls were placed
inside the cell, one side of which was covered with a thin copper tape
providing an
effectively conducting side wall boundary condition. TWs formed on the
insulating
side wall at $R_{cs}$ but were suppressed on the conducting wall. The direction
of
the TWs is $\vec{\Omega} \times \vec{n}$ where $\vec{\Omega}$ is the rotation
vector
and $\vec{n}$ the the normal
vector of the wall into the fluid. A straight wall can thus be considered as
the
side wall for an effectively infinite convection cell. Kuo and Cross have
recently
calculated the properties of this straight wall state using free-free top and
bottom boundary conditions and rigid boundary conditions at the lateral
wall.\cite{KC93}

\begin{figure}
\centerline{\psfig{file=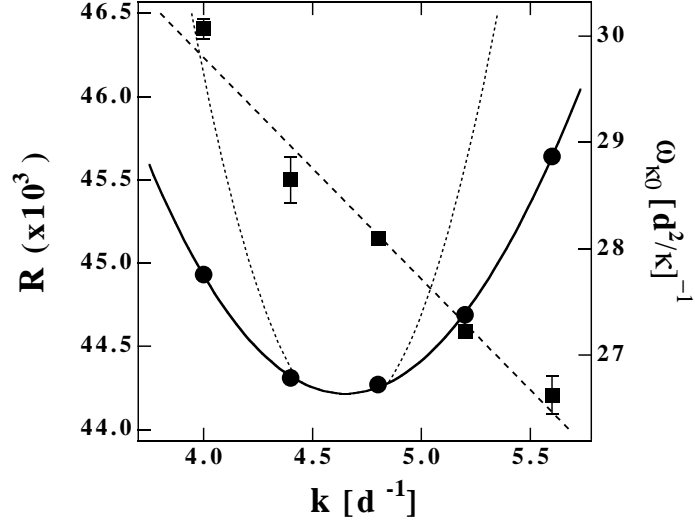,height=2.9in}}
\caption{Marginal stability boundary in parameter space of R and azimuthal
wavenumber k ($\bullet$) and
$\omega_{\kappa 0}$ vs k (\fbox{\protect\rule[0.35ex]{0.09em}{0cm}})
for $\Omega = 544$ with $\Gamma = 2.5$. The solid line
is a nonlinear third-order polynomial fit to the marginal stability data, the
short-dashed line is the finite-size Eckhaus boundary, and the long-dashed
line is a linear fit to the frequencies.}
\label{fig:stability}
\end{figure}

\begin{figure}
\centerline{\psfig{file=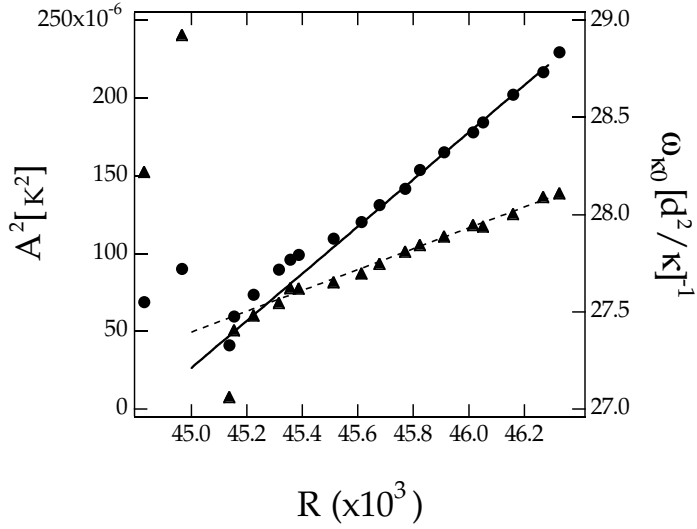,height=2.9in}}
\caption{Temperature amplitude squared A$^2$ ($\bullet$)
and frequency $\omega_\kappa 0$ ($\triangle$) vs $R$
for the transition from mode 13 to mode 12.}
\label{fig:amp}
\end{figure}

The amplitude of the wall convection state is concentrated near the side
wall and behaves rather like a quasi-one-dimensional periodic pattern along the
circumference. At $\Omega = 544$, we prepared several states with different
mode
numbers by jumping to a high $\Delta T$ and quickly dropping back to just above
the critical $\Delta T$. We then measured the amplitude of the traveling wall
state and determined the marginal stability boundary by extrapolating the
squared
amplitude versus $R$ to zero. Figure \ref{fig:stability} shows this marginal
stability
boundary and the
corresponding precession frequency measured for $\Omega = 544$ in the $\Gamma =
2.5$
cell. From fits to the data we obtain $R_c$ = 44217, $k_c$ = 4.65, and
$\omega_{\kappa 0c}$ =
28.3. The actual switching of mode number toward the critical one occurred at
finite
amplitudes, suggesting an instability other than marginal instability. To
identify this
instability, we prepared an $m = 13$ state at $R$ = $4.63 \ times 10^4$
and ramped down to $R_{cs}(m=12)$
over about $80$ hours making measurements at intervals of $\delta R
\approx 0.002$.
We define $\epsilon$ in terms of $R_c = 44217$ for the marginal stability
curve.
At $R = 4.547 \times 10^4$,  $\epsilon = 0.0283$ with $R_c = 44217$ for the
marginal
curve, traveling wave's amplitude began to deviate from a linear dependence on
$R$ (see Figure \ref{fig:amp}). The transition to the critical mode number $m_c
= 12$ took place $27$ hours ($32\tau_\kappa$) later. The extrapolated marginal
stability of the $m = 13$ mode is at $R = 4.469 \times 10^4$ ($\epsilon =
0.0107$).
Fig. \ref{fig:amp} also shows the measured precession frequencies scaled by
$\tau_\kappa$.

	Because of the quasi one-dimensionality of the traveling wall states,
it should be described by a 1-D complex Ginzburg-Landau (CGL) equation,
$$\tau_0 \left ({\partial A\over\partial t} + s_0 {\partial A\over\partial x}
\right ) = \epsilon (1+ic_0) A + \xi_0^2 (1+ic_1){\partial^2 A\over\partial
x^2} -
g_0(1+ic_3)|A|^2 A.$$ Since the precession is opposite to the rotation
direction
frequencies are negative.
We have measured all the coefficients in the CGL equation at $\Omega = 544$,
and
they are $c_0 = 0.6, \ c_1 = 0.6, \ -0.01 < c_3 < 0 , \ \tau_0 = 0.027,
\ s_0 = 2.1, \ \xi_0 = 0.192$, and $g_0 = 0.8$, scaling time by
$d^2/\kappa$ and  length by $d$. In this equation,
$\epsilon$ is defined relative to $R_c$ = 44217 and the modulation wavenumber
relative to $k_c$ =4.65, both
obtained from the fit to the marginal stability curve,
Fig.~\ref{fig:stability}.
Using these definitions we obtain the linear coefficients as follows:
$\xi_0$ is the quadratic curvature of the marginal stability curve,
Fig.~\ref{fig:stability} ; $\tau_0$ is obtained from transient measurements of
the
oscillation amplitude; $s_0$ is the slope of the frequency at the critical
wavenumber, Fig.~\ref{fig:amp}; the combination $c_1-c_0$ is determined from
the curvature of the frequency dispersion, Fig.~\ref{fig:amp}; the combination
$c_3-c_0$ comes from the variation of frequency with $\epsilon$. To resolve the
remaining degree of freedom we measure the transient frequency response. To
double check our result that $c_3$ is small we use a modulation technique in
which we sinusoidally vary the bottom plate temperature with a
period larger than 5.2$\tau_\kappa$. This method allows us to probe
several parameters at once and to develop better statistics than is
possible for the single transient measurements.
These results are consistent with the values
obtained through the other static and dynamic measurements.
The nonlinear coefficient is obtained by normalizing the local probe
temperature
with the heat transport using $Nu-1 = |A|^2 R/R_c$ and with an area correction
since the traveling-wave state is confined to a region near the wall
of order the depth of the fluid (we take a radial width $d$ so $g_0 =
(dNu/d\epsilon)(\Gamma^2/(2\Gamma -1))$.
We can compare our linear coefficients, $c_0, c_1, \tau_0, s_0$, and $\xi_0$
with
predictions of the linear theory for our particular experimental
conditions\cite{MNK92}
which yield $c_0 = 0.7, \ c_1 = 0.8,  \ \tau_0 = 0.022, \ s_0 =
1.8,$ and $\xi_0 = 0.21,$ in reasonable agreement with the experimental
values. As in the experiments the CGL equation is not exact so there are finite
size uncertanties in the numerical coefficients. Infinite-system, nonlinear
calculations
\cite{KC93} for a traveling-wave convection state on a flat wall for
$\Omega$=540 and with
free-free top-bottom boundary conditions yield  $c_0 = 0.8, \ c_1 = 0.45, \
c_3 = 0.32, \ \tau_0 = 0.022,
\ s_0 = 2.1, \xi_0 = 0.22$, and $g_0 = 0.6$.
Here again the agreement between theory and experiment is quite
satisfactory except with regard to the nonlinear coefficient c$_3$.
For more quantitative comparisons the calculations need to incorporate
rigid-rigid boundaries and the experimental geometry needs to be changed to
accomodate many more periods, thereby better approximating the large-system
limit.
{}From the experimental values one obtains $1+c_1 c_3>0$ (the Newell criterion)
indicating that the uniform traveling wave pattern is Benjamin-Feir stable. The
stability of the traveling wave is bounded by the Eckhaus stable wavenumber
band\cite{JPBCRK92}
$$\epsilon_E = {2(1+c_3^2) + 1+ c_1 c_3 \over 1 + c_1 c_3}\xi_0^2 (k-k_c)^2.$$
Outside this
stability boundary, the pattern is unstable to long wavelength perturbations
and will
undergo a subcritical bifurcation to a new pattern with a wavenumber within the
stability boundary. Given the experimentally obtained values for $c_1$ and
$c_3$, we
find that $\epsilon_E = 3\xi_0^2(k-k_c)^2 = 3 \epsilon_M$.
In Fig.~\ref{fig:stability}, we plot this Eckhaus boundary with finite size
corrections\cite{Tu90}. This boundary agrees well
with the experimental evidence that a transition took place around $\epsilon =
0.0283$ when $\epsilon_M = 0.0107$. Since the transition was from $m=13$ to
$m=12$,
both the amplitude and the precession frequency contained progressively more
$m=12$
component, thus becoming larger than if the state was purely mode 13, see Fig.
\ref{fig:amp}. We observed this variation during the transition.  Since our
cell is
very uniform and the quasi one dimensional wall convection structure is
strictly
periodic, there is virtually no long wavelength perturbation to force the
instability
to occur. This gives rise to long-lived ``metastable'' states (Zhong {\it et
al}) when
certain wall states have non-critical mode numbers that are outside the Eckhaus
stability
boundary.

\subsection{Bulk convection state}

As $\Delta T$ increased above the first bifurcation to the wall state, we
observed a
second  transition in the Nusselt number measurements. Accompanied with this
rise in
heat transport, we also observed convective motion throughout the cell (the
``bulk''
state) and an increase in the thermal noise in the bottom plate which was
supplied
with a constant heat power. This indicates that the second transition was to
time
dependent convection, which was confirmed by analyzing time sequences of many
shadowgraph images of the convection pattern.

\begin{figure}
\centerline{\psfig{file=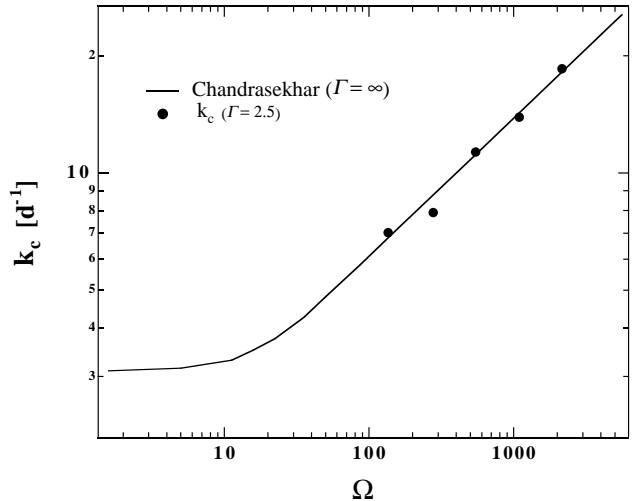,height=2.9in}}
\caption{The critical bulk wavenumber $k_{cb}$ vs $\Omega$ for the bulk
state obtained from shadowgraph images. The solid line is from
Chandrasekhar\protect{\cite{Ch61}}.}
\label{fig:wavec}
\end{figure}

To characterize the bulk convection state, we measured its critical Rayleigh
number $R_{cb}$ and critical wavenumber $k_{cb}$ for $\Omega < 2145$.
The critical Rayleigh numbers, determined from heat transport measurements,
are shown in Figure \ref{fig:pdiag}. Also plotted as a solid line is
Chandrasekhar's
linear stability analysis result for a laterally infinite system\cite{Ch61}.
The
critical wavenumbers obtained from  recorded shadowgraph images are shown in
Figure \ref{fig:wavec}. They also agree well with the results of Chandrasekhar,
shown
as the solid line.
{}From each shadowgraph image, we extracted their central portions,
applied 2-D Fourier transformation with band pass frequency filtering and
summed
up the total power. Just above the onset where the shadowgraph image intensity
variation is almost linear in the temperature gradient, the total power in the
spectrum is $\sum_i |A_i|^2$ if the convection structures are composed of
rolls of  $A_i exp(i {\vec k _i} \cdot {\vec x})$. In Figure \ref{fig:images}
(a), (b)
and (c),
\begin{figure}
\vskip 4.0in
%\centerline{\psfig{file=images.ps,height=4.0in}}
\caption{Shadowgraph images of convection pattern (a) below ($R = 4.87 \times
10^5$,
$\epsilon$ = -0.062), (b) nearly at
($R = 5.11 \ times 10^5$, $\epsilon$ = -0.016), and (c) above
($R = 5.38 \times 10^5$, $\epsilon$ = 0.037) the onset of
bulk convection, and (d) the effective Nusselt number Q/\'Q ($\bullet$), ie.
the ratio of
the total heat conducted in the convecting state
divided by the heat conducted by the side-wall state alone,
and the squared amplitude (\fbox{\protect\rule[0.35ex]{0.09em}{0cm}}) vs
$\epsilon$.
All the data are for $\Omega$ = 2145 and the $\epsilon$ values for the patterns
are
labeled in (d).}
\label{fig:images}
\end{figure}
we show the background-divided and digitally-enhanced images of the convection
patterns with $\Omega$ = 2145
for values of $R$ just below, nearly at and slightly above $R_{cb}$,
which was determined from the change in slope of the Nusselt number.
The $\epsilon$ values in the figure are defined relative to $R_{cb}$.
To more easily
compare the bulk amplitude obtained
from spectral analysis and heat transport data, we define an
effective Nusselt number as the ratio of heat conducted by both convection
states
(bulk and side-wall) divided by the heat conducted by the side-wall state
alone.
Figure \ref{fig:images} (d) shows this effective
Nusselt number and $\sum_i |A_i|^2$ versus $\epsilon$ for $\Omega = 2145$.
The linear relation
between the total squared amplitude and $\epsilon$ confirms that the
bifurcation
is forward. We conclude that the bulk convection state is the one given by
Chandrasekhar's linear stability analysis\cite{Ch61}; because there is no
appreciable
amplitude of the side-wall state in the central portion the bulk state can grow
from
small amplitude despite being a second transition.

Figure \ref{fig:pdiag} shows that the bifurcation lines for the side wall state
and the
bulk convection state cross each other when $60<\Omega<70$ for $\Gamma = 2.5$.
For
$\Omega$ less than the intercept value, bulk convection sets in first and wall
convection begins later. The bulk and wall convections are spatially separated,
interacting only weakly. The cross-over point for the bifurcation lines is
therefore
not a co-dimension two point. A further point about this regime at small
$\Omega$ is that the small aspect ratio of the container may become important.
Qualitatively nothing changes but for the quantitative comparison with theory
these
finite size effects are probably significant. In particular, the bulk
state is the initial bifurcation and its presence
might strongly influence the side-wall bifurcation which now happens at higher
R; the high value, relative to theory,
of the side-wall onset for $\Omega = 16.93$ suggests such an effect.
The onset of the bulk state is also affected by the small aspect ratio when
$\Omega$ is small as evidenced by the small upward shift of $R_{cb}$ consistent
with finite-size effects in non-rotating convection\cite{CS70,SM75}.
Both of these shifts are visible in Fig.~\ref{fig:pdiag}.

There are several additional interesting aspects of the results obtained from
shadowgraph images. First, there is a range of $R$ around $R_{cb}$ where the
convection pattern in the middle region consists of concentric rings. As is
known\cite{ACHS81}, concentric ring patterns in
Rayleigh-B\'enard convection can be induced by thermal forcing from the side
wall
owing to the mismatch of thermal conductivities between the side wall and
working
fluid. In our case, the effective side wall for the bulk convection is the wall
state, which is convecting rather vigorously at the second bifurcation point.
For $\Omega = 1090$ and near the onset of bulk convection ($R/R_{cs} = 2.246$),
$Nu = 1.54$. The active convection area is at most $50\%$ of the cell, implying
an effective thermal  conductivity of the wall state at least twice that of the
conducting water in the middle. The concentric rings eventually break up when
$\epsilon$ is increased and give rise to cellular structures.

The total squared amplitude ($\sum_i |A_i|^2$), shown in Figure
\ref{fig:images} (d),
has a linear dependence on $R$. However, the extrapolation of this linearity to
the base line yields a Rayleigh number slightly less than $R_{cb}$, typically
on the
order of 0.2\%.
Judging from the rounded corner of the $Nu$ versus $\epsilon$ curve near
$\epsilon
= \epsilon_2$, this shift is caused by the appearance of the concentric ring
pattern.

A time sequence of the images taken at different points in parameter
space reveals the source of the time dependence in the bulk convection state.
For $\Omega$ as high as $2145$, we observed a K\"uppers-Lortz
roll-orientation switching and/or turbine like rotation of convection cells.
The
dynamics stemming from the K\"uppers-Lortz instability are aperiodic, which
gives
rise to the increased thermal noise in the bottom plate. This aspect of the
result will be presented in greater detail elsewhere\cite{NE92}.

\section{Conclusion}

In this paper, we presented experimental results for the initial
bifurcations in rotating Rayleigh-B\'enard convection in small aspect ratio
cylindrical cells. We confirmed a forward bifurcation from
conduction to convection near rigid walls. Our experimental data agree
beautifully with the theoretical results for the rotation dependence of the
critical
traveling wave frequency and mode number. We confirmed the
prediction that the critical frequency goes linearly to zero as $\Omega$
approaches zero. We also conjecture from the experiment that for insulating
walls,
the critical mode number will reach an asymptotic finite value as
$\Omega$ increases. We have shown
for the second bifurcation that the convection states are very similar to the
ones that are linearly unstable in the infinite system. Because of their
spatial
separation, the wall and bulk convection states interact only weakly in a
narrow annular
region between the two states.
As aspect ratio increases, the separation becomes more distinct and the
wall state becomes relatively more confined to the side wall. Since the wall
state is quasi one dimensional, its contribution to the dynamics and transport
properties of rotating thermal convection should diminish in the limit of large
aspect ratio. The presence of such ``surface'' modes will, however, subtly
affect
aspects of the bulk nonlinear dynamics, as is evident by its forcing of the
concentric ring pattern near the onset of bulk convection.

The authors wish to thank H. F. Goldstein, E. Knobloch, I. Mercader and M. Net
and E. Kuo and M. Cross
for allowing us to use their theoretical results prior to publication. We
gratefully acknowledge conversations with M. Cross, L. Kramer, W. van Saarloos,
L.
Tuckerman, and K.
Babcock.  One of us (RE) would like to acknowledge the Institute for
Theoretical Physics
at UC Santa Barbara for participation in the program ``Spatially-Extended
Nonequilibrium Systems.'' This research was supported by the U.S. Department of
Energy.

\end{document}